# Bright and Purcell-enhanced single photon emission from a silicon G center

*Kyu-Young Kim,* [‖,†] *Chang-Min Lee,*[⊥,†] *Amirehsan Boreiri,*[⊥] *Purbita Purkayastha,*[⊥] *Fariba Islam,*[⊥] *Samuel Harper,*[⊥] *Je-Hyung Kim,* [‖] *and Edo Waks*[⊥,#,*]

‖ Department of Physics, Ulsan National Institute of Science and Technology, Ulsan 44919, Republic of Korea

⊥ Institute for Research in Electronics and Applied Physics, University of Maryland, College Park, Maryland 20742, USA

# Joint Quantum Institute, University of Maryland, College Park, Maryland 20742, USA

Abstract

Silicon G centers show significant promise as single photon sources in a scalable silicon platform. But these color centers have large non-radiative decay and a low Debye-Waller factor, limiting their usability in quantum applications. In this work, we demonstrate bright Purcell-enhanced emission from a silicon G center by coupling it to a nanophotonic cavity. The nanobeam cavity enhances the spontaneous emission rate of a single G center by a factor of 6, corresponding to a Purcell factor greater than 31 when accounting for decay into the phonon sideband. We obtain a spontaneous emission rate of 0.97 ns, which is the fastest single photon emission rate reported in silicon. With this radiative enhancement, we achieve an order of magnitude improvement in emitter brightness compared to previously reported values. These results pave the way for scalable quantum light sources on a silicon photonic chip.

Silicon color centers have recently emerged as promising quantum emitters for scalable quantum silicon photonic devices [1–5]. Various color centers in silicon have been isolated as single quantum emitters, including W centers [6], T centers [7–11], and G centers [12–19]. Among these choices, G centers are particularly appealing as quantum light sources. The G center emission lies in the telecom O-band, making it compatible with fiber optic networks. These emitters also feature a fast excited state lifetime of ~ 5 ns, making them compelling candidates for bright single photon sources[15–20]. However, G centers suffer from a low Debye-Waller factor of 15% [3,16,20], indicating that the majority of the photons emit into a broad incoherent sideband rather than coherent zero-phonon line, as well as low quantum efficiency due to non-radiative decay[17,19,21].

Nanophotonic cavities provide a solution to the low efficiency of the G center by enhancing their spontaneous emission rate via the Purcell effect. This strategy was effectively employed in other color centers including T centers[9–11] and W centers[22]. Cavity-enhanced emission was also reported in G* centers that emit at a wavelength that is similar to the G center[14]. But all of these color centers exhibit significantly longer radiative decay rates than the G center, with the G* center and the W center having the shortest excited state lifetime of ~ 30 ns. Thus, they require significantly larger Purcell enhancement to achieve similar brightness to the G center emission. By directly enhancing the spontaneous emission rate of the G center, one could potentially engineer an ultra-bright single photon emitter in silicon. Indeed, recent work demonstrated significant brightness enhancement of a G center coupled to a photonic crystal cavity[16]. But no enhancement in the radiative decay rate was observed due to the presence of competing non-radiative decay processes.

In this work, we report bright Purcell-enhanced single photon emission from a cavity-coupled silicon G center. We employ a nanophotonic cavity that both enables efficient photon collection into an optical fiber and enhances spontaneous emission. Time-resolved photoluminescence measurement reveals a six-fold lifetime enhancement of a cavity-coupled G center, corresponding to the Purcell factor of 31. We achieve a Purcell-enhanced spontaneous emission rate of 0.97 ns, the fastest reported rate from a single photon emitter in silicon. By coupling the cavity emission to an adiabatically tapered waveguide we also attain high photon collection efficiency, resulting in a source to fiber brightness of 14%. These results constitute a step towards bright scalable single photon sources in silicon photonics.

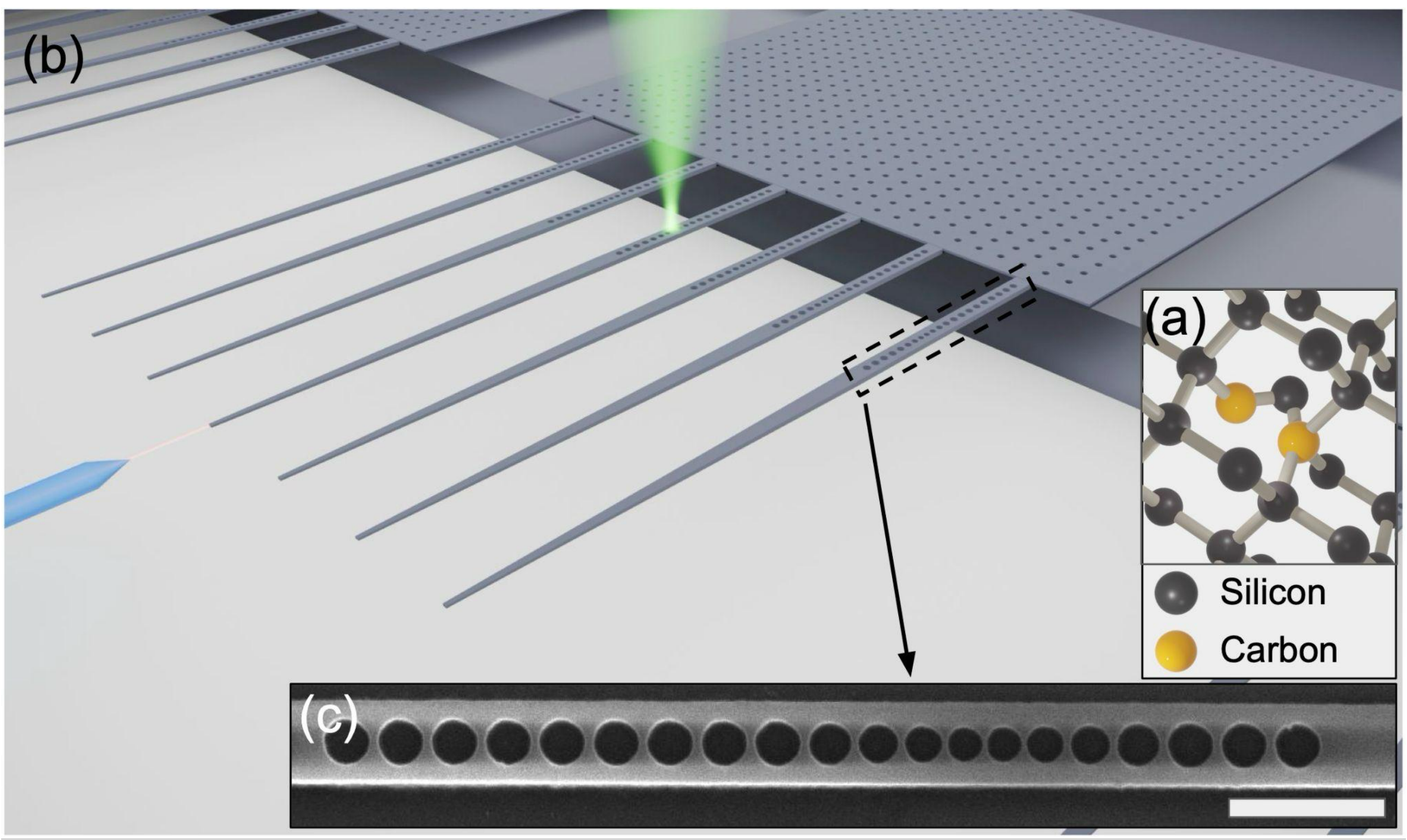

Figure 1. (a) Atomic structure of a silicon G center. (b) A schematic image of a nanobeam cavity array and a lensed fiber. (c) Scanning electron microscope image of the cavity part of the fabricated nanobeam cavity. Scale bar is 1 μm.

Figure 1(a) shows the atomic structure of the G center in silicon, consisting of two substitutional carbon atoms and an interstitial silicon. It emits in the telecom O-band with a zero-phonon line at 1278 nm. To create the G centers, we utilized a commercial silicon-on-insulator wafer with a device layer thickness of 220 ± 10 nm. We implanted Carbon-12 atoms at 35 KeV with the fluence of $7\times10^{12}$ ions/cm$^2$ (CuttingEdge Ions) followed by rapid thermal annealing at 1000°C for 20 seconds. The implantation condition targets ion implantation depth of ~110 nm (Ref. [15]). We note that the G centers are randomly distributed horizontally.

To enhance the emission of the G center, we couple it to a nanobeam photonic crystal cavity. Figure 1(b) illustrates the cavity structure. The cavity comprises a one-dimensional photonic crystal with a defect engineered by tapering the hole radius and the periodicity. The details of the cavity design were reported in previous work[9]. In this work we use the same design but modify the beam width, the hole radius, and the periodicity to shift the cavity resonance to 1277 nm so that it is aligned with the G center emission. We use an asymmetric cavity with 8 holes on the right and 13 holes on the left to ensure that the majority of the cavity field emits towards a tapered waveguide that is mode-matched to a lensed fiber. Supporting information section 1 provides additional details about the cavity design, including simulated mode volume of $0.26\left(\frac{\lambda}{n}\right)^3$, where $\lambda$ is the resonant wavelength of the cavity mode in free space, and $n$ is the refractive index of the cavity material. We fabricated the designed nanobeam cavities using standard e-beam lithography, chemical wet etching, and transfer printing using a Polydimethylsiloxane (PDMS) micro-stamp (see Ref. [8] for the fabrication recipe details). Figure 1(c) shows a scanning electron microscope image of a fabricated device.

We performed all measurements at 8 K in a closed cycle cryostat system. The vacuum chamber contained a fiber probe station for lensed fiber coupling. We tuned the cavity resonance by depositing a thin layer of nitrogen on the sample and heating up the sample to remove the layer [23,24]. A pulsed diode laser with a wavelength of 780 nm was focused at the nanobeam cavity to excite the G centers. We collected the G center emission through the lensed fiber, and measured it using a spectrometer system or a fiber spectral filter (transmission window of 0.1 nm) followed by superconducting nanowire single photon detectors. Supporting Information section 2 contains more details on the measurement setup.

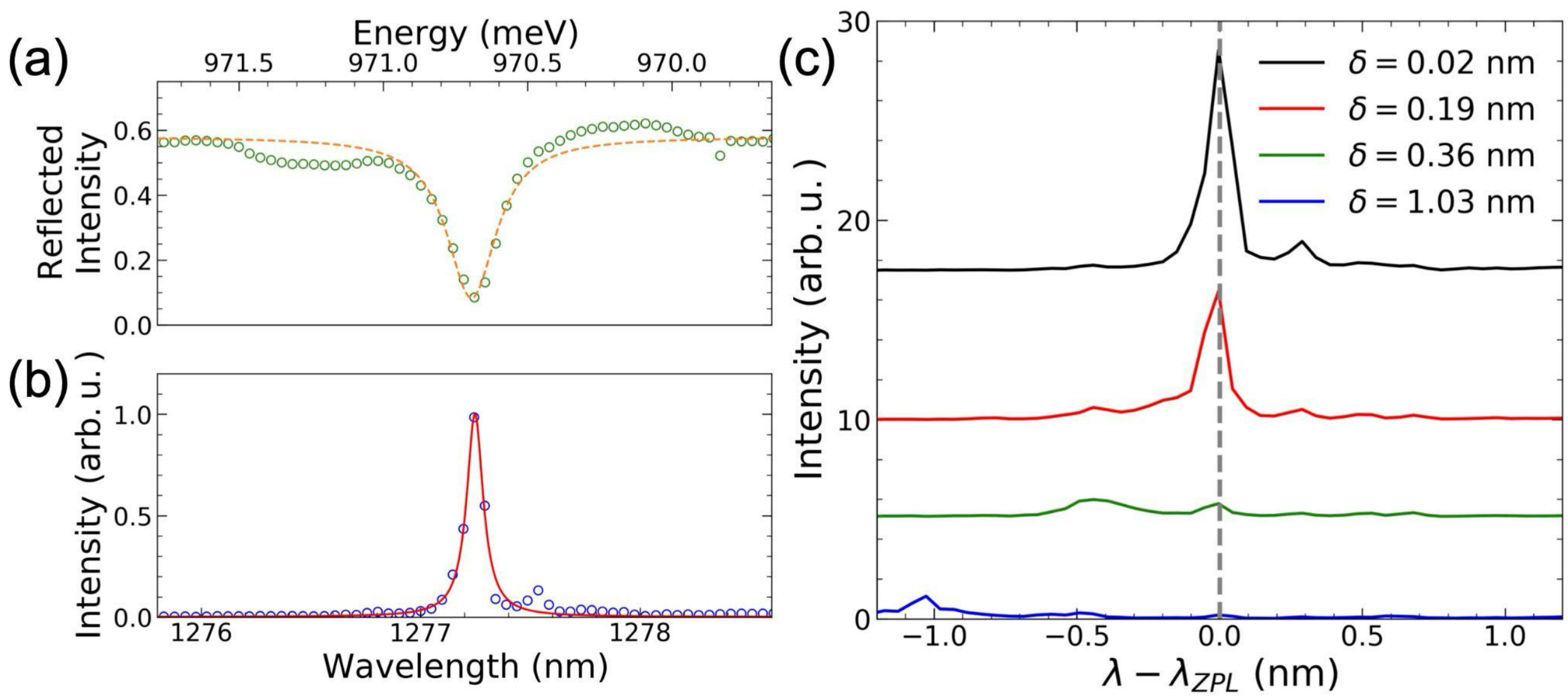

Figure 2. (a) Measured reflectivity spectrum of a nanobeam cavity. The green circles are measured data and the yellow dashed line is a Lorentzian fit. (b) Above-band photoluminescence spectrum of a cavity-coupled G center. Red solid line corresponds to a Lorentzian fit of the peak. (c) Photoluminescence spectra of the cavity-coupled G center at different detuning $\delta = \lambda_{ZPL} - \lambda_{cav}$. Gray dashed line indicates the center wavelength of the zero-phonon line of the G center.

We first characterize the nanobeam cavity mode via reflectivity measurement through the lensed fiber (see Supporting Information section 2 for the details). Figure 2(a) shows a reflectivity spectrum of the cavity, obtained with a broadband tungsten-halogen lamp input. We first tune the cavity to the expected G center resonant frequency and then probe the reflection spectrum through the lensed fiber. The reflectivity exhibits a clear dip at 1277.23 nm corresponding to the cavity mode. The dip approaches near zero on resonance, indicating that this cavity is operating close to critical coupling. A Lorentzian fit of the data (yellow dashed line in Figure 2(a)) reveals a quality factor of 4600 for the cavity mode. We independently measure the coupling efficiency from nanobeam to fiber by tuning the laser off-resonance from the cavity so that it acts as a mirror, and measuring the reflected power. From these measurements we measure the coupling efficiency between the nanobeam and the lensed fiber to be 75% (Supporting Information section 2).

To generate photons from the cavity-coupled G center, we excited the cavity from the top using a high numerical aperture microscope objective and collected the emission from the lensed fiber. We used a 780 nm pulsed laser with a tunable repetition rate of 10-40 MHz. Figure 2(b) shows the measured spectrum using a pump power of 2.0 μW and repetition rate of 40 MHz. The spectrum exhibits a sharp peak at 1277.25 nm that corresponds to the zero-phonon line of silicon G centers. From a Lorentzian fit of this peak (red solid line), we measured the linewidth of 0.091 nm, significantly narrower than the cavity spectral bandwidth. In addition, photoluminescence from the G center ensemble shows a local vibration mode (E-line) at 1381 nm, which is another fingerprint of G centers in silicon (Supporting Information section 5).

Figure 2(c) plots the emission spectrum of the G center as a function of cavity detuning. We plot all spectra as a function of the detuning $\delta = \lambda_{ZPL} - \lambda_{cav}$, where $\lambda_{ZPL}$ is the wavelength of the G center zero-phonon line and $\lambda_{cav}$ is the wavelength of the cavity mode. At each detuning condition, we obtained the $\lambda_{cav}$ separately from cavity reflectivity measurements. We observe a clear and rapid reduction of the emission intensity as we detune the cavity, indicating that the cavity strongly enhances the brightness of the zero-phonon line.

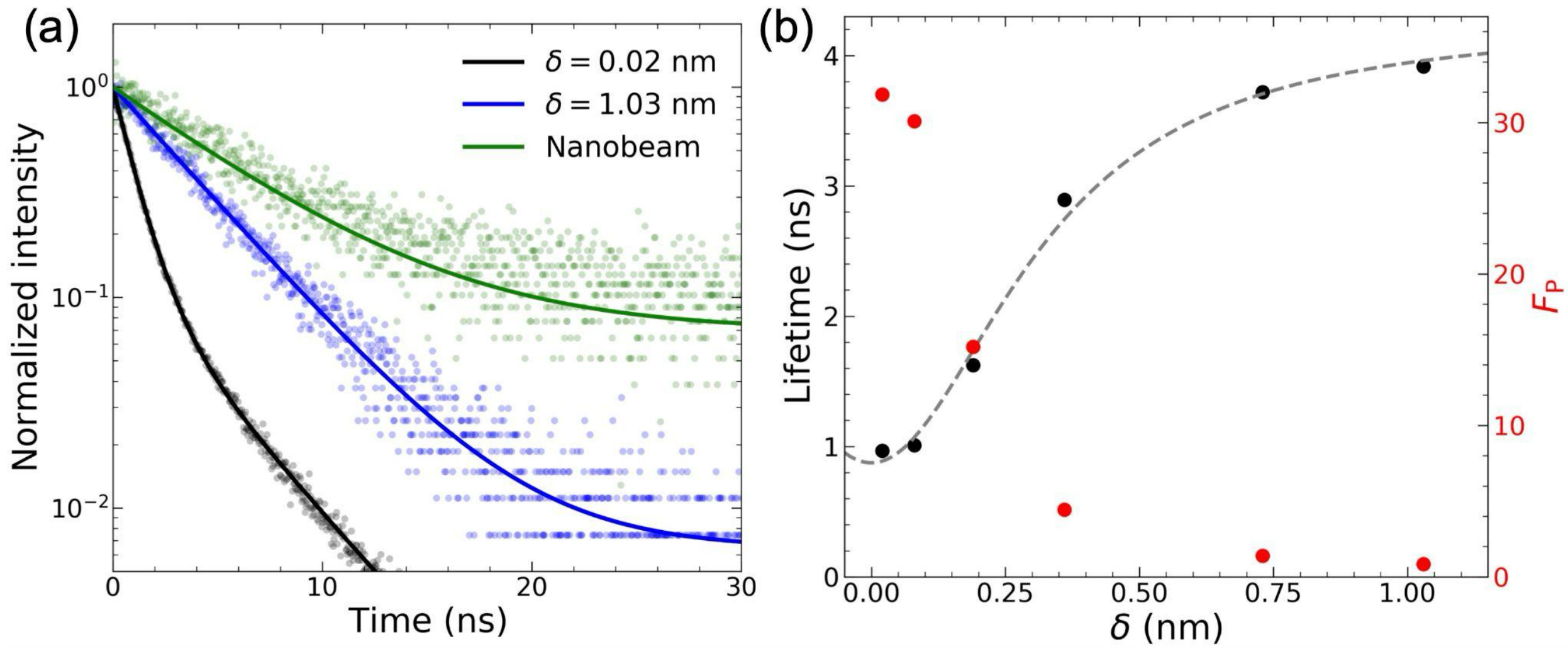

Figure 3. (a) Time-resolved photoluminescence using above-band pulsed excitation with a repetition rate of 10 MHz. Black (blue) circles are the measured data at detuning of $\delta = 0.02$ nm ($\delta = 1.03$ nm). Green circles are obtained from a G center located in the nanobeam outside the cavity region. Solid curves correspond to exponential decay function fit. (b) Reduced lifetime (black dots) and the estimated Purcell factor $F_P$ (red dots) as a function of detuning $\delta$. Black dashed line is a Lorentzian fit of the reduced lifetime.

To further validate that the cavity is enhancing the emission rate, we performed time-resolved lifetime measurements. Figure 3(a) shows the time-resolved emission decay for two different tunings. When the G center is on resonance with the cavity mode ($\delta = 0.02$ nm; black circles), the lifetime data fit well to a bi-exponential decay function (black solid line) with a fast decay time of $\tau_{on} = 0.97 \pm 0.01$ ns and a slow decay time of $3.75 \pm 0.04$ ns. The fast decay corresponds to the Purcell-enhanced zero-phonon line of the G center. We attribute the slow decay to the emissions from other G centers not coupled to the cavity that lie in the same wavelength window. As a reference, we also measured the lifetime of the G centers in the nanobeam outside of the cavity region (green circles). The lifetime decays fits well a single exponential function with decay time of $\tau_0 = 5.97 \pm 0.01$ ns. By comparing the on-resonance decay time $\tau_{on}$ to the out-of-cavity decay time $\tau_0$, we determine a six-fold lifetime enhancement based on the nanobeam cavity.

Figure 3(a) also shows the time-resolved lifetime measurement for when we detune the G center from the cavity $\delta = 1.03$ nm (blue circles). In this case the emitter decay time increases to $3.92 \pm 0.02$ ns, demonstrating that the decay rate enhancement is induced by the cavity. We performed the same lifetime measurements at different detunings (Figure 3(b), black dots) and the data fit well to the Lorentzian function (black dashed line). From the Lorentzian fit, we obtained an off-resonance lifetime $\tau_{off} = 4.27$ ns. This value is shorter than the out-of-cavity lifetime $\tau_0$. We attribute this reduction in lifetime to additional non-radiative decay which may be induced by proximity to etched sidewalls within the cavity structure.

To estimate the Purcell factor, we must account for the non-radiative decay of the emitter due to the phonon sideband and additional mechanisms. We can estimate the Purcell factor $F_P$ using the equation

$$F_P = \tau_0 \left(\frac{1}{\tau_{on}} - \frac{1}{\tau_{off}}\right) \frac{1}{F_{DW}} \frac{1}{\varepsilon_{rad}}, \tag{1}$$

where $F_{DW}$ is the Debye-Waller factor without any cavity, and $\varepsilon_{rad}$ is quantum efficiency of the G center (see Supporting Information section 3 for the derivation of the Purcell factor). Using the G center Debye-Waller factor[3,16,20] of 15% and the measured lifetimes, we obtain a lower bound of the Purcell factor $F_P \geq 31$. The Purcell factor will increase even further if we factor in the quantum efficiency, which ranges 1%-10% from previous reports[17,19,21]. We note that the estimated maximum Purcell factor based on the quality factor and the mode volume of the nanobeam cavity is 1340, and mismatch of position and dipole orientation between the G center and the cavity mode may have degraded the Purcell factor from the estimated value (Supporting information Section 3). The strong Purcell effect results in the zero phonon line emission becoming dominant over other decay processes. Subsequently, we estimate source brightness (defined as the probability of the G center emitting a zero-phonon line photon) to be $1 - \tau_{on}/\tau_{off} = 0.77$ (see Supporting Information section 3).

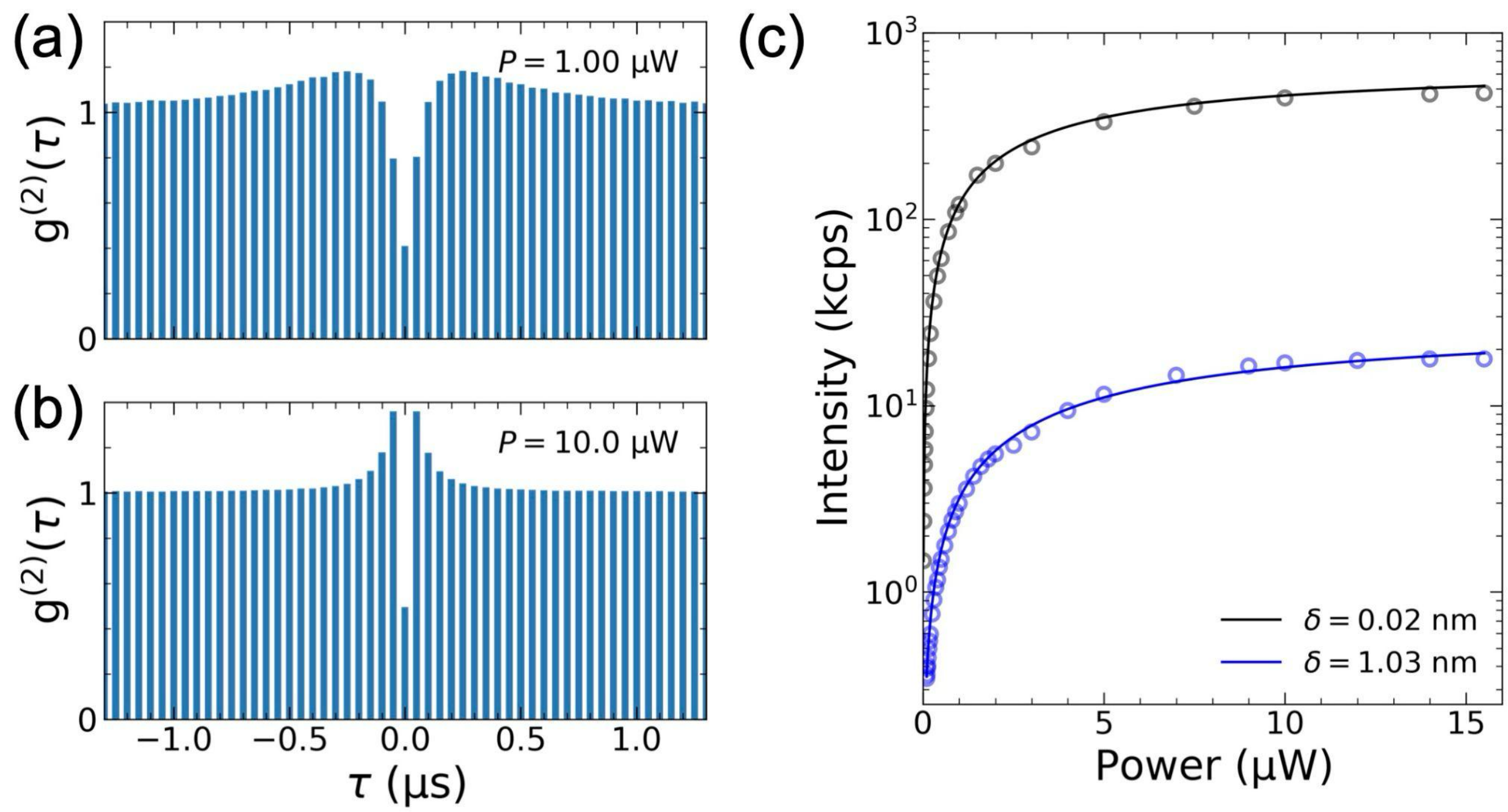

Figure 4. (a,b) Second order correlation measurement of the zero-phonon line using a pulsed excitation with a repetition rate of 20 MHz, at average power of (a) 1.0 μW and (b) 10 μW. (c) Pump power dependence of detected count rates at $\delta = 0.02$ nm (black circles) and $\delta = 1.03$ nm (blue circles). Solid lines are fit with a two-level atom saturation model.

We next performed second-order autocorrelation measurements to verify that this emission corresponds to a single G center. We excited the G center using a pulsed 780 nm laser with a repetition rate of 20 MHz. Figure 4(a) and 4(b) show measured autocorrelations at an excitation power of 1 μW and 10 μW respectively. The bar graph plots the total counts integrated over a 50 ns window around each peak of the correlation measurement. At an average pump power of $P =$ 1.0 μW, we obtained $g^{(2)}(0) = 0.408 \pm 0.004$ (Figure 4(a)). We followed the algorithm by T. Wohland *et al* to calculate the second order correlations[25]. This value is less than 0.5, which confirms single photon emission from the G center emitter. The $g^{(2)}(0)$ value is limited by contributions from other G centers emission. As we increase the pump power up to $P = 10$ μW, the $g^{(2)}(0)$ value increases to $0.494 \pm 0.002$. Both Figure 4(a) and 4(b) show additional slow dynamics which is known as coming from the existence of metastable states [14,16,17].

To investigate the source brightness of the single G center, we measured single photon count rates as a function of pump power at a detuning of $\delta = 0.02$ nm and $1.03$ nm (Figure 4(c)). We utilized a 780 nm pulsed laser with repetition rate of 10 MHz. For both on-resonance ($\delta = 0.02$ nm, black circles) and off-resonance ($\delta = 1.03$ nm, blue circles), the count rates fit well with a two-level atom model $I(P) = I_{sat}P/(P + P_{sat})$ where $I_{sat}$ is saturation intensity and $P_{sat}$ is saturation pump power[26]. When we tuned the cavity mode onto the G center emission ($\delta = 0.02$ nm), we obtained $I_{sat} = 669$ kcps (kilocounts per second) from the fitting. The count rates in Figure 4(c) contain both single photon states from the G center and background from other off-resonant emitters. They therefore overestimate the source brightness. To exclude multiphoton events originating from background emitters, we calculate the single photon count rate using the formula[27] $I_{corr} = I_{sat} \times \sqrt{1 - g^{(2)}(0)}$. From this equation, the corrected single photon count rate is $I_{corr}$ = 476 kcps. Dividing this count rate by the 10 MHz repetition rate of the pulsed laser, we obtain a total end-to-end brightness of 4.76%. This total brightness is at least one order of magnitude higher than the previous work, which reported count rates of 10 kcps (Ref. [14–17]).

To determine the source efficiency, we perform a photon budget to account for the losses of all optical components and collection optics. These losses include detection efficiency of the detectors (90%), transmission of fiber components (78%), transmission of fiber tunable filter (47%), lensed fiber coupling efficiency (75%), and nanobeam-to-fiber coupling efficiency (50%) (See Supporting Information section 4). From these losses we estimated the fiber-coupled brightness of 14% and the source brightness of 38%. We attribute the discrepancy between this source brightness and the estimated source brightness from the lifetime measurements (Figure 3) to potential error in the nanobeam-to-fiber coupling efficiency, polarization of the single photon input of the single photon detectors, and transmission of fiber components. We performed the same measurement at a detuning of $\delta = 1.03$ nm and achieved $I_{corr} = 25.4$ kcps. This value is 19 times smaller than the on-resonance photon count rate, indicating that the Purcell effect induced significant intensity enhancement of the G center emission.

In conclusion, we demonstrated nanobeam cavity enhanced emission from the zero-phonon line of a single G center. We obtained a six-fold lifetime reduction corresponding to the Purcell factor lower bound of 31, resulting in a radiative lifetime of 0.97 ns. This is the fastest emission rate reported in a silicon single photon source. Our current experiments were conducted at 8 K, where thermally induced dephasing may reduce the coherence times. Reducing the sample temperature could improve the coherence time, opening up the possibility to achieve high photon indistinguishability. In addition, engineering nanocavities with higher quality factors will increase the G center emission rate further, approaching the lifetime-limited photon emission regime. These results represent a significant advancement towards utilizing silicon G centers as efficient and indistinguishable single photon sources in silicon.

AUTHOR INFORMATION

**Corresponding Author**

Edo Waks (edowaks@umd.edu)

**Author Contribution**

†K.-Y.K. and C.-M.L. contributed equally to this work.

**Acknowledgements**

The authors would like to acknowledge financial support from the National Science Foundation (grants #OMA1936314, #OMA2120757, #ECCS2423788, and #ECCS1933546), the U.S. Department of Defense, the National Research Foundation of Korea grant funded by MSIT (RS-2024-00438839, RS-2024-00442762), and the education and training program of the Quantum Information Research Support Center, funded by MSIT of the Korean government (No.2021M3H3A1036573).

# Supporting Information for

# Bright and Purcell-enhanced single photon emission from a silicon G center

*Kyu-Young Kim,* [∥,†] *Chang-Min Lee,*[⊥,†] *Amirehsan Boreiri,*[⊥] *Purbita Purkayastha,*[⊥] *Fariba Islam,*[⊥] *Samuel Harper,*[⊥] *Je-Hyung Kim,* [∥] *and Edo Waks*[⊥,#,*]

∥ Department of Physics, Ulsan National Institute of Science and Technology, Ulsan 44919, Republic of Korea

⊥ Institute for Research in Electronics and Applied Physics, University of Maryland, College Park, Maryland 20742, USA

# Joint Quantum Institute, University of Maryland, College Park, Maryland 20742, USA

**Section 1. Dimensions of the nanobeam cavity design and finite-difference time-domain simulation of the cavity mode**

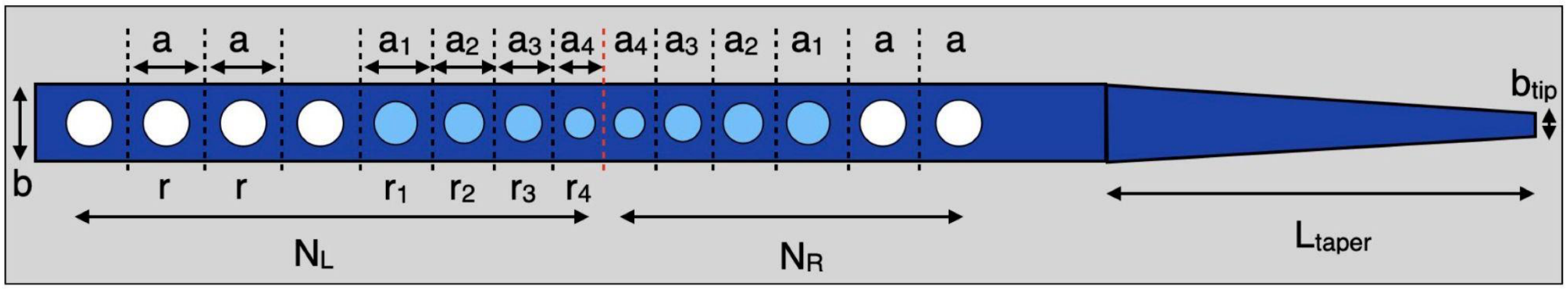

Figure S1. Schematic of the nanocavity design. The photonic crystal cavity is formed from a four-hole linear taper defect in a periodic hole array. The parameters of the cavity are the number of holes on the left ($N_L$) and right ($N_R$), the periodicity of the photonic crystal hole array (a) and taper ($a_i$), the hole radii of the mirror (r) and taper ($r_i$), the width of the nanobeam (b), the tip width of the nanobeam taper ($b_{tip}$), and the length of the nanobeam taper ($L_{taper}$).

The nanobeam cavity comprises a one-dimensional air-clad silicon nanobeam photonic crystal cavity with an adiabatic waveguide taper on one side, which allows for outcoupling to a lensed fiber. The photonic crystal cavity is composed of an array of air holes, with the cavity region formed by linearly tapering both the hole radii and periodicity. The number of holes to the left ($N_L$) and right ($N_R$) can be individually selected to control the in-plane coupling in either direction. In our device, we employed a four-hole taper. We optimized the cavity using commercial finite-difference time-domain simulation software to achieve a high quality factor mode at the G center resonance. From these simulations, we determined an optimal value for the lattice constant a of 350 nm and a hole radius r of 115 nm, with the hole radii in the cavity region ($r_1$ - $r_4$) tapered from 107.5 nm to 85 nm, while the hole periodicities ($a_1$ - $a_4$) were tapered from 325 to 250 nm. Our design used an $N_R$ of 8 and an $N_L$ of 13 to ensure the majority of the light is emitted to the tapered region. For the adiabatic taper end, we utilized b = 520 nm, $b_{tip}$ = 120 nm and $L_{taper}$ = 12 µm.

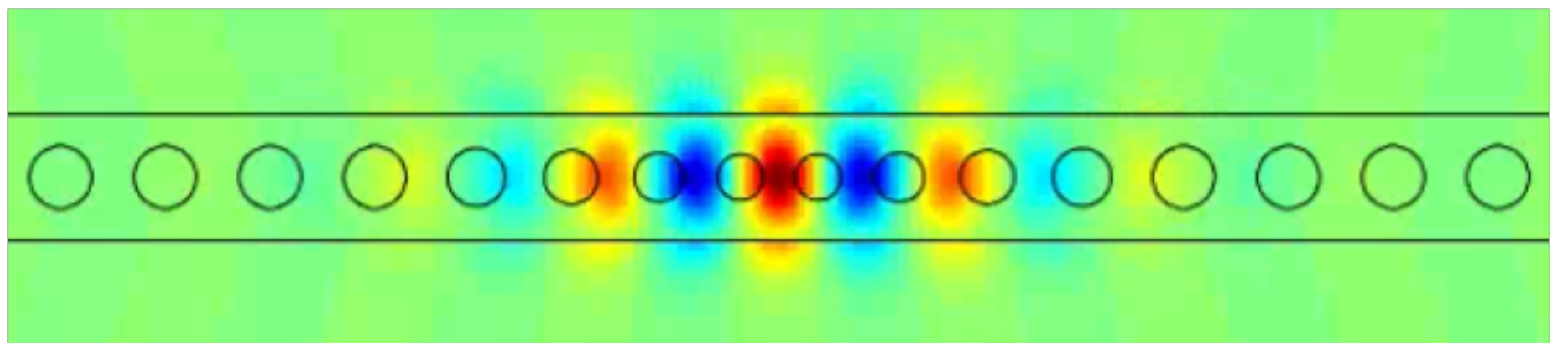

Figure S2. Electric field distribution of the cavity mode from the finite-difference time-domain simulation.

To characterize a cavity mode of the designed nanobeam cavity, we performed finite-difference time-domain simulation using a commercial software (Ansys Lumerical FDTD). Figure S2 shows electric field distribution of the cavity mode. From this simulation, we achieved mode volume of the cavity mode $0.26\left(\frac{\lambda}{n}\right)^3$, where $\lambda$ is the resonant wavelength of the cavity mode in free space, and $n$ is refractive index of the cavity material.

**Section 2. Optical measurement setup for cavity reflectivity and photoluminescence of the G center**

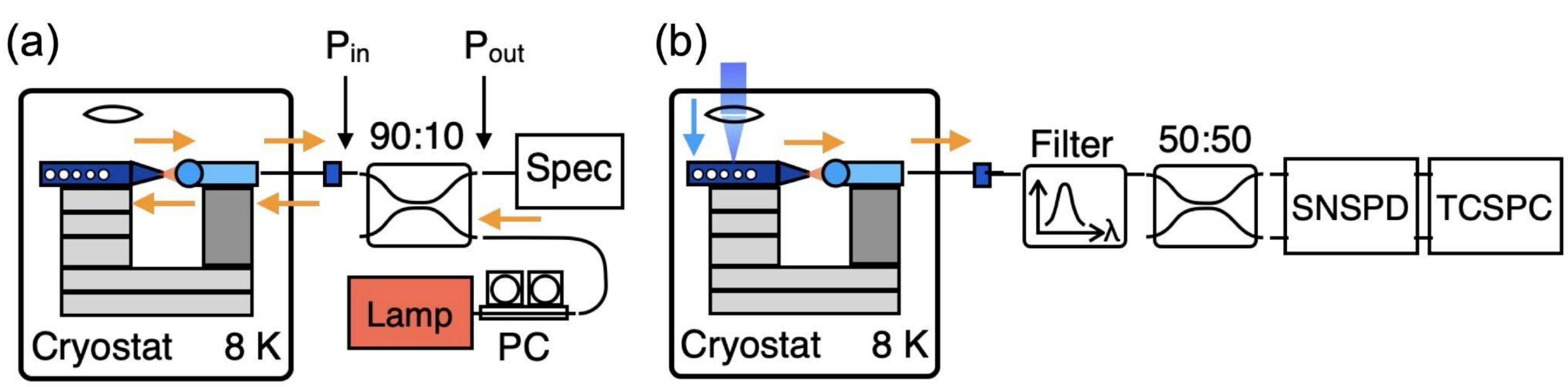

Figure S3. (a) Measurement setup for cavity reflectivity. PC: polarization controller, Spec: spectrometer. (b) Measurement setup for single photon counting. SNSPD: superconducting nanowire single photon detectors, TCSPC: time-correlated single photon counter.

Figure S3(a) depicts the reflectivity setup to measure broad-band reflectivity spectrum of the nanobeam cavity (Figure 2(a) in the main manuscript). We sent a broad-band tungsten-halogen lamp through the lensed fiber and the reflected light was measured by a spectrometer and charge-coupled device array system. Figure S3(b) displays the photon counting setup to measure single photon count rate, lifetime, and second-order autocorrelation. In this case, we focused an excitation laser beam on the nanobeam from the top optical window. The pulsed excitation laser has a wavelength of 780 nm and repetition rates of 10-40 MHz range. We collected the emission through the lensed fiber and filtered it with a fiber-type tunable spectral filter with a transmission window of 0.1 nm. For the photon counting, the filtered signal was measured using superconducting nanowire single photon detectors with detection efficiency of 90% and a time-correlated single-photon counter.

We estimated the coupling efficiency between the nanobeam and the lensed fiber $\eta_{ficoup}$ by measuring reflectivity. We assumed that the coupling-in and coupling-out efficiencies are the same. The input and output power were measured at the $P_{in}$ and $P_{out}$ positions, respectively (Figure S3(a)). Assuming that the reflectivity of the photonic crystal mirror is unity, the off-resonance reflectivity is given by $R = P_{out}/P_{in} = \eta_{ficoup}{}^2 T_{fc}$, where $T_{fc}$ is the transmission of the 90:10 fiber coupler (89%). To assess the $\eta_{ficoup}$, we replaced the tungsten-halogen lamp in Figure S3(a) with a narrow-band laser at 1280 nm and measured the reflectivity using a power meter. From this measurement we obtained $R = 50\%$, corresponding to $\eta_{ficoup} = 75\%$.

**Section 3. Purcell factor and quantum efficiency**

Let $\gamma_0$ be the decay rate of the G center in silicon without any cavity. Then

$$\gamma_0 = \gamma_z + \gamma_{SB} + \gamma_{nr}, \tag{1}$$

where $\gamma_z$ is the decay rate into zero-phonon line, $\gamma_{SB}$ is the decay rate into the sideband, and $\gamma_{nr}$ is the non-radiative decay rate. Additionally, we have

$$\gamma_{rad} = \gamma_z + \gamma_{SB} = \varepsilon_{rad}\gamma_0, \tag{2}$$

and

$$\gamma_z = F_{DW}\gamma_{rad} = F_{DW}\varepsilon_{rad}\gamma_0, \tag{3}$$

where $\varepsilon_{rad}$ is the quantum efficiency and $F_{DW}$ is the Debye-Waller factor without any cavity.

Now, let $\gamma_{cav}$ be the decay rate of the G center inside a cavity. Then

$$\gamma_{cav}(\omega) = F_P \frac{(\Delta\omega)^2}{4(\omega-\omega_0)^2+(\Delta\omega)^2}\gamma_z + \gamma_{SB} + \gamma_{nr}', \tag{4}$$

where $\gamma_{nr}'$ is the non-radiative decay rate inside a cavity, $\omega$ is the frequency of the zero-phonon line of the G center, $\omega_0$ is the frequency of the cavity mode, $\Delta\omega$ is the linewidth of the cavity mode, and $F_P$ is the Purcell factor when $\omega = \omega_0$. Here, when we tune the cavity far from the G center, then the first term goes to zero. For both off-resonance and on-resonance, we have:

$$\gamma_{off} = \gamma_{SB} + \gamma_{nr}', \tag{5}$$

and

$$\gamma_{on} = F_P \gamma_z + \gamma_{SB} + \gamma_{nr}' = F_P \gamma_z + \gamma_{off}, \tag{6}$$

where the $\gamma_{off}$ ($\gamma_{on}$) is the decay rate when the cavity is off-resonance (on-resonance).

Using equations (3) and (6), we find an expression for the Purcell factor:

$$F_P = \frac{\gamma_{on} - \gamma_{off}}{\gamma_z} = \frac{\gamma_{on} - \gamma_{off}}{\gamma_0} \frac{1}{F_{DW}} \frac{1}{\varepsilon_{rad}} = \tau_0 \left( \frac{1}{\tau_{on}} - \frac{1}{\tau_{off}} \right) \frac{1}{F_{DW}} \frac{1}{\varepsilon_{rad}}, \tag{7}$$

where $\tau_{on} = 1/\gamma_{on}$ ($\tau_{off} = 1/\gamma_{off}$) is the decay time when the cavity is on-resonance (off-resonance), and $\tau_0 = 1/\gamma_0$ is the decay time of the G center without any cavity. In the main manuscript, we measured $\tau_{on} = 0.97\ ns$, $\tau_{off} = 4.27\ ns$, and $\tau_0 = 5.97\ ns$. From the literature, the Debye-Waller factor $F_{DW}$ was estimated to be 0.15. Using these values, we obtain a lower bound for the Purcell factor:

$$F_P = 31.2 \times \frac{1}{\varepsilon_{rad}} > 31.2. \tag{8}$$

We can also estimate source brightness from the measured as:

$$Brightness = F_P\gamma_z/\gamma_{on} = (\gamma_{on} - \gamma_{off})/\gamma_{on} = 1 - \tau_{on}/\tau_{off} = 0.77. \quad (9)$$

When the emitter's dipole orientation aligns with the cavity mode profile and the emitter's location is at the maximum field position of the cavity mode, we calculate the maximum Purcell factor using this formula $F_P = \frac{3}{4\pi^2}\frac{Q}{V}\left(\frac{\lambda}{n}\right)^3$, where $Q$ is quality factor of the cavity, $V$ is mode volume of the cavity mode, $\lambda$ is resonant wavelength of the cavity mode in free space, and $n$ is refractive index of the cavity material. Using the quality factor of 4600 (Figure 2(a) in the main manuscript) and the mode volume of $0.26\left(\frac{\lambda}{n}\right)^3$ (Supporting Information section 1), we estimate the maximum Purcell factor of 1340.

**Section 4. Photon budget analysis**

To estimate the brightness of this single photon source, we perform photon budget analysis. From the pump power dependence of the single photon count rate and $g^{(2)}(0)$ value displayed in the main manuscript, we measured the $g^{(2)}(0)$-corrected count rate $I_{corr}$ = 476 kcps. Using the repetition rate of the excitation laser $R_{rep}$= 10 MHz, we obtained an end-to-end brightness of 4.76%. Between the single photon detectors and the lensed fiber, there exist three efficiency/transmission to consider: SNSPD detection efficiency $\eta_{SNSPD}$ of 90%, transmission of fiber cables and fiber beam splitter $\eta_{BS}$ of 78%, and transmission of fiber tunable filter $\eta_{filt}$ of 47%. Consequently, we calculated the fiber-coupled brightness of the single photons $(I_{corr}/R_{rep})/(\eta_{SNSPD}\eta_{BS}\eta_{filt})$ = 14%.

To estimate the source brightness, there are two more efficiencies to consider. As described in Supporting Information section 2, we obtained lensed fiber coupling efficiency $\eta_{ficoup}$ = 75%. In the main manuscript, Figure 2(a) explains the cavity-waveguide coupling efficiency $\eta_{WG}$ of 50%. Using these two values, we estimated the source brightness as $(I_{corr}/R_{rep})/(\eta_{SNSPD}\eta_{BS}\eta_{filt}\eta_{ficoup}\eta_{WG})$ = 38%.

**Section 5. G center ensemble photoluminescence**

To investigate general properties of the defect centers, we fabricated nanobeam waveguide structures using the same implanted silicon wafer described in the main manuscript. We then measured the photoluminescence spectrum of the nanobeam waveguide from 1260 to 1500 nm wavelength range using the same measurement setup (Figure S4(a)). We observed a strong peak at 1277 nm corresponding to the zero-phonon line of the silicon G centers. Furthermore, we found another characteristic E-line emission at 1381 nm, which is attributed to a local vibration mode of the G center[1]. The presence of the E-line further supports the conclusion that the collected emission is from silicon G centers.

Figure S4(b) shows the zero-phonon line of the G center ensemble at high excitation laser power. We fitted the spectrum data using Gaussian function and achieved a spectral linewidth of 1.28 nm (red solid line). This linewidth is comparable to other reported inhomogeneous broadening values in the range of 1.1-1.3 nm for similar G centers[2,3].

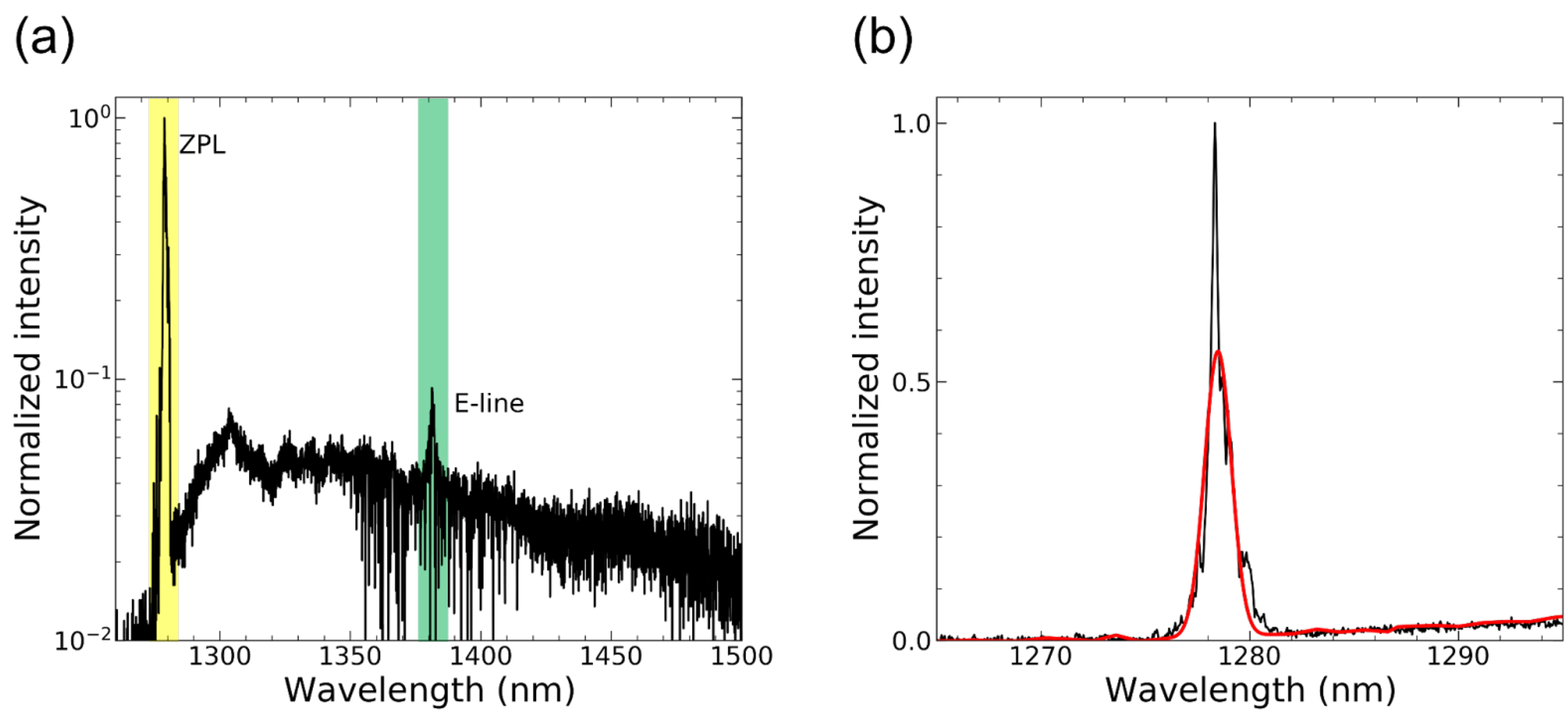

Figure S4. Photoluminescence spectra of G center ensemble in a nanobeam waveguide. (a) Wide range photoluminescence spectrum in log scale. Yellow and green boxes are the zero-phonon line (1277 nm) and E-line (1381 nm) of the G center ensemble. (b) Photoluminescence spectrum at high excitation laser power (black line) and its Gaussian fitting (red line) with a linewidth of 1.28 nm.